\documentclass[a4paper,11pt]{article}
\usepackage{amsmath, amssymb, graphicx, physics, bm}
\usepackage[a4paper, margin=1in]{geometry}

\usepackage{jheppub, appendix, mathrsfs} 
\usepackage{lineno}

\title{\boldmath Superadditivity of Krylov Complexity for Tensor Products}

\author{Jeff Murugan and Hendrik J. R. van Zyl}
\affiliation{The Laboratory for Quantum Gravity \& Strings,\\
Department of Mathematics and Applied Mathematics,\\
University of Cape Town, Private Bag, Rondebosch, 7701,\\
South Africa}

\emailAdd{jeff.murugan@uct.ac.za}
\emailAdd{hjrvanzyl@gmail.com}

\abstract{We study Krylov complexity for quantum systems whose Hamiltonians factorise as tensor products. We prove that complexity is superadditive under tensor products, $C_{12}\ge C_1+C_2$, and identify a positive operator that quantifies the resulting excess complexity. The underlying mechanism is made transparent by introducing a Krylov graph representation in which tensor products generate a higher-dimensional lattice whose diagonal shells encode operator growth and binomial path multiplicities. In the continuum limit, Krylov dynamics reduces to diffusion on this graph, with superadditivity arising from geometric broadening across shells. Explicit examples illustrate how deviations from synchronous evolution generate bounded, oscillatory excess complexity.
}

\begin{document}
\maketitle
\flushbottom

\section{Introduction}
\label{sec:intro}
The growth of quantum complexity has emerged as a central theme in quantum many-body physics, quantum information, and quantum gravity. Among the various notions of complexity, Krylov (or spread) complexity \cite{Parker:2018yvk, Balasubramanian:2022tpr, Nandy:2024evd, Rabinovici:2025otw, Baiguera:2025dkc} has proven particularly useful as a diagnostic of operator growth \cite{Parker:2018yvk,Roberts:2018mnp,Jian:2020qpp}, quantum chaos \cite{ Avdoshkin:2022xuw,  Balasubramanian:2022tpr, Hashimoto:2023swv, Erdmenger:2023wjg, Alishahiha:2024vbf, Baggioli:2024wbz}, and thermalisation \cite{Alishahiha:2024rwm}, with direct connections to Lanczos dynamics \cite{Dymarsky:2019elm, Rabinovici:2021qqt, Nandy:2024evd}, spectral statistics \cite{Bhattacharyya:2023grv}, and emergent geometric descriptions of quantum evolution \cite{Caputa:2021sib, Chattopadhyay:2023fob}. In this framework, a Hamiltonian $H$ and a reference state $|\phi\rangle$ generate a Krylov basis via the Lanczos algorithm \cite{Lanczos1950AnIM}, producing an orthonormal sequence of states $\{|K_n\rangle\}$ that captures how repeated action of $H$ explores Hilbert space. The associated spread complexity measures how probability weight distributes across this basis and has been shown to encode rich dynamical information, including early-time growth, late-time saturation, and universal scaling regimes \cite{Camargo:2024deu}.\\

\noindent
A basic but surprisingly subtle question concerns the behaviour of Krylov complexity under composition of systems. Given two subsystems with Hamiltonians $H_1$ and $H_2$, one may ask how the complexity of the combined system relates to the complexities of the individual subsystems. For many notions of complexity, additivity under tensor products is either assumed or built in by construction \cite{Nielsen:2012yss}. For Krylov complexity, however, the situation is far from obvious: the Krylov basis of the combined Hamiltonian
$H = H_1 \otimes I + I \otimes H_2$
is generically not the tensor product of the individual Krylov bases, and the Lanczos orthogonalisation introduces nontrivial interference effects.\\

\noindent
In this work we show that spread complexity is generically superadditive under tensor products. More precisely, for any target state supported on the Krylov subspace of the combined system, the corresponding complexity satisfies $C_{12} \ge C_1 + C_2$,
where $C_1$ and $C_2$ are the complexities associated with the subsystems. We construct an explicit positive semi-definite operator that measures the resulting excess complexity, and we give a fully analytic proof of this inequality. Importantly, we identify a sharp criterion for saturation: the bound is saturated if and only if the two subsystems evolve synchronously, in the sense that the raising operators appearing in the Lanczos decomposition align. Beyond establishing superadditivity, we characterise the structure of the excess complexity operator in detail. Its diagonal elements measure the deviation between Krylov indices and tensor-product indices, while its off-diagonal elements encode the variance of this mismatch. This allows us to interpret excess complexity as arising from the appearance of orthogonal states within fixed tensor-product shells that are accessed only at higher Krylov order.\\

\noindent
We illustrate these ideas explicitly using tensor products of $su(2)$ Hamiltonians, where the Krylov basis is known in closed form \cite{Caputa:2021sib, Chattopadhyay:2023fob}. These examples make clear how excess complexity first appears at higher Krylov levels, explain the observed early-time scaling behaviour of the time-evolved reference state, and demonstrate how synchrony suppresses excess complexity. A complementary and unifying perspective emerges from a graphical viewpoint. We show that the tensor-product Krylov basis naturally defines a two-dimensional Krylov graph, generalising the familiar one-dimensional Krylov chain. The graph has a square-lattice structure organised into diagonal shells of constant tensor-product index, with a combinatorial structure governed by binomial coefficients. In a suitable continuum limit, the dynamics of Krylov amplitudes is described by a two-dimensional advection–diffusion equation. This hydrodynamic description provides a geometric explanation of superadditivity and connects excess complexity to the curvature of probability flow on the Krylov graph.\\

\noindent
Our results generalise straightforwardly to multipartite tensor products, where the Krylov graph becomes higher-dimensional and the diagonal shells form simplices governed by multinomial combinatorics. More broadly, this work clarifies how operator growth, combinatorics, and orthogonalisation conspire to produce genuinely new complexity in composite quantum systems.

\section{Krylov Complexity of Tensor Products}

Consider two subsystems with Hamiltonians $H_1$ and $H_2$ and reference states $\ket{\phi_1}$ and $\ket{\phi_2}$. Together, the Hamiltonian and reference state pairs generate \cite{Altman:2014fhm,Caputa:2021sib} the Krylov bases $\{\ket{K_{1,n}}, n \in 0,1,2,\cdots, N_1\}$ and \hbox{$\{\ket{K_{2,n}}, n \in 0,1,2,\cdots, N_2\}$} respectively. Note that we use $N_1$ and $N_2$ for the index of the most complex state in the respective subspaces and that the Krylov dimensions are $N_1+1$ and $N_2+1$ respectively.  There are several algortihms \cite{Lanczos1950AnIM, Muck:2022xfc} that may be used to compute these bases though these are not particularly relevant for the present discussion.  We will only make use of the fact \cite{Beetar:2023mfn} that
\begin{eqnarray}
(H_1)^m|\phi_1\rangle & = & \sum_{j_1=0}^m \alpha_{m;j_1}|K_{1,j_1}\rangle  \nonumber \\
(H_2)^n|\phi_2\rangle & = & \sum_{j_2=0}^n \beta_{n;j_2}|K_{2,j_2}\rangle   \label{KforSubsystem}
\end{eqnarray}
i.e. the $m$'th Krylov basis vector only starts appearing after $m$ applications of the Hamiltonian on the reference state.
Now consider the combined Hamiltonian
\begin{equation}
H = H_1 \otimes I + I \otimes H_2
\end{equation}
which acts on the tensor product reference state $\ket{\phi_r} = \ket{\phi_1} \otimes \ket{\phi_2}$, generating a joint Krylov basis $\left\{ |K_n\rangle    \right\}$ consisting of $N+1$ elements. The $n$-th Krylov vector of the combined system can be extracted from the states
\begin{equation}\label{tensor-expansion}
(H_1 \otimes I + I \otimes H_2)^n \ket{\phi_r} = \sum_{k=0}^n \binom{n}{k} (H_1)^k \ket{\phi_1} \otimes (H_2)^{n-k} \ket{\phi_2}.   
\end{equation}
by a Gram-Schmidt process.  Orthogonalization of these combinations defines a new Krylov basis whose level index $n$ corresponds asymptotically to the total ``degree'' $m+n$ in powers of $H_1$ and $H_2$.  One may expect the dominant contribution at large $n$ to be governed by binomial weights $\binom{n}{k}$, yielding a structure reminiscent of Pascal's triangle.  We will return to this idea later, but for now we will keep the discussion general without any approximations.  \\ \\
Some aspects of the Krylov basis for the combined system are useful to take note of.  Firstly, the Krylov dimension has an upper bound due to the tensor product structure.  Secondly, the $n$'th Krylov basis element will always contain the states
$$ |n; a\rangle \equiv |K_{1,n-a}\rangle \otimes |K_{2,a}\rangle   $$
as a component.  These considerations imply that 
$$ N_1 + N_2 \leq N \leq (N_1+1)(N_2+1)-1   $$
This inequality saturates the lower bound if the two subsystems evolve in synchrony, a concept that we will still unpack in detail.  The Krylov dimension of the combined system can thus be much larger than the Krylov dimension of the subsystems and states that are far more complex may be explored.  We will expand on intuitive ways to understand this in due course.  \\
\noindent
Finally, defining the complexity operators 
\begin{eqnarray}
\hat{C}_{12} & = & \sum_{n=0}^N n \ket{K_n}\bra{K_n}     \nonumber   \\
\hat{C}_1 & = & \sum_{n=0}^{N_1} n |K_{1,n}\rangle \langle K_{1,n}|    \nonumber \\  
\hat{C}_2 & = & \sum_{n=0}^{N_2} n |K_{2,n}\rangle \langle K_{2,n}|    \nonumber 
\end{eqnarray}
we find the inequality
\begin{equation}
\langle \psi| \hat{C}_{12} |\psi\rangle \geq  \langle \psi_1| \hat{C}_1 |\psi_1\rangle +   \langle \psi_2 | \hat{C}_2   | \psi_2\rangle,
\end{equation}
where $|\psi\rangle = |\psi_1\rangle \otimes |\psi_2\rangle$  is \textbf{any} target state supported on the Krylov basis $|K_n\rangle$.  In other words, spread complexity is superadditive when considering tensor products.   The inequality is saturated when the two subsystems evolve in synchrony.  
To prove this inequality we first note that the operators $\hat{C}_1$ and $\hat{C}_2$ can be expressed in terms of the tensor product basis as
\begin{eqnarray}
\hat{C}_1 & = & \sum_{m,n} m |m+n; n\rangle \langle m+n; n|   \nonumber   \\
\hat{C}_2 & = & \sum_{m,n} n |m+n; n\rangle \langle m+n; n|    
\end{eqnarray}
by using the fact that the resolution of the identity in terms of the Krylov bases (acting on their respective subspaces).  The superadditivity of spread complexity is thus the statement that the matrix 
\begin{equation}
\langle K_m | \hat{\Delta}_C|K_n\rangle \equiv \langle K_m | \hat{C}_{12} - \hat{C}_1 - \hat{C}_2 |K_n\rangle    \label{excessMat}
\end{equation}
is positive definite.  Being a strictly non-negative quantity it quantifies the ``excess'' complexity of the tensor product for a given target state.    

\subsection{Proof of superadditivity}

When considering numerical examples, the easiest way to confirm that (\ref{excessMat}) is positive semi-definite is to compute its eigenvalues.  In this section, however, we will prove it analytically for which we have found thatthe simplest way is to verify that $\langle \Psi | \hat{\Delta}_C |\Psi\rangle \geq 0$ for all states $|\Psi\rangle$.  The proof of this statement rests largely on a single, simple observation.  The expressions (\ref{KforSubsystem}) and (\ref{tensor-expansion}) imply that
\begin{eqnarray}
H^n |K_{1,0}\rangle \otimes |K_{2,0}\rangle = \sum_{k=0}^n \binom{n}{k} \sum_{j_1 = 0}^{n-k} \sum_{j_2 = 0}^{k} \alpha_{n-k, j_1} \beta_{k,j_2} |j_1+j_2; j_2\rangle    \nonumber
\end{eqnarray}
From this it immediately follows that 
$$  \langle K_n | m+m'; m'\rangle  = 0 \ \ \ \textnormal{if} \ \ \ n > m+m'$$
In other words, states that require a combined $n$ applications of $H_1$ and $H_2$ to start appearing in the Lanczos algorithm can thus only start appearing at $n$'th order.  Intuitively, this already suggests that the excess complexity will be strictly non-negative.   One would expect that generating states by actions of the combined Hamiltonian $H_1 + H_2$ and superposition will require more effort than generating them with both the Hamltonians $H_1 + H_2$, $H_1 - H_2$ and superposition.  \\ \\
To make the proof precise we will now assume\footnote{This assumption breaks if $N \neq N_1 N_2$ since, in this case, there are states in the tensor product basis that are not supported on the Krylov basis.  This is easy to overcome, however.  If the Krylov basis only covers part of the tensor product basis i.e. $N_K < N_1 N_2$ we can define $|K_{N_K+1}\rangle$ to be any state orthogonal to the  existing Krylov basis vectors and continue the Lanczos algorithm.  We may repeat this several times until we have a total of $N_1 N_2$ orthogonal vectors.  Using this augmented basis the statement of super-additivity holds true for every state in the tensor product space and not just those supported on the Krylov subspace. } that every state in the tensor product space can be written as 
\begin{equation}
|n; b\rangle = \sum_{q=n}^{N} \alpha_{n,b}^q |K_q\rangle    \label{tensorInTermsofK}
\end{equation}
We may suppress the summation bounds if we define $\alpha_{n, b}^q = 0$ for $n > q$ and $q > N$.  Orthogonality of the states imply that the coefficients satisfy
\begin{equation}
\sum_{q} \alpha_{m;a}^q (\alpha_{p;b}^q)^* = \delta_{mp}\delta_{ab}
\end{equation}
and, being eigenstates of both $\hat{C}_1$ and $\hat{C}_2$, the states $|n; m\rangle$ satisfy
\begin{eqnarray}
\langle m; a | \hat{C}_1 + \hat{C}_2 | n; b\rangle & = &  \delta_{m,n} \delta_{a,b} \ m.
\end{eqnarray}
In what follows we will only need to make use of (\ref{tensorInTermsofK}) to prove that $\langle \Psi | \hat{\Delta}_C |\Psi\rangle \geq 0$ and thus that $\hat{\Delta}_C$ is a positive semi-definite operator.  It is possible to go directly to the most general state $|\Psi\rangle$, but it is instructive to do so gradually.  

\subsubsection{Diagonal entries}

The simplest example of a target state is to just take the vector $|m; a\rangle$.   The diagonal elements of $\hat{\Delta}_C$ are given by
\begin{eqnarray}
\langle m; a| \hat{\Delta}_C |m; a\rangle & = & \sum_{n=m}^N n |\alpha_{m;a}^n|^2 - m   \nonumber  \\
& = &  \sum_{n=m}^N n |\alpha_{m;a}^n|^2 - m  \sum_{n=m}^N |\alpha_{m;a}^n|^2    \nonumber \\
& = & \sum_{k=1}^{N-m} k |\alpha_{m;a}^{m+k}|^2.  
\end{eqnarray}
which is clearly non-negative.

\subsubsection{Eigenstates of $\hat{C}_1 + \hat{C}_2$}

Next we consider a general eigenstate of $\hat{C}_1 + \hat{C}_2$
$$ |\psi_m\rangle = \sum_{a}  \sigma_a^m |m; a\rangle  $$
The operator  $\hat{C}_1 + \hat{C}_2$ is block diagonal and the value $m$ may be any integer value from $0$ up to $N_1 + N_2$.  In addition to the diagonal elements of the previous subsection we will also need the off-diagonal elements within the same block for $\hat{C}_{1} + \hat{C}_{2}$.  These are given by 
\begin{eqnarray}
\langle m; a| \hat{\Delta}_C |m; b\rangle & = &  \sum_{ q=m }^N q (\alpha_{m; a}^q)^*(\alpha_{m; b}^q)    \nonumber \\
& = & \sum_{ q=m }^N q (\alpha_{m; a}^q)^*(\alpha_{m; b}^q)  - m \sum_{ q=m }^N (\alpha_{m; a}^q)^*(\alpha_{m; b}^q)     \nonumber \\
& = & \sum_{k=1}^{N-m} k (\alpha_{m; a}^{m+k})^* (\alpha_{m; b}^{m+k})
\end{eqnarray}
We note that this formula also captures the diagonal elements when $a = b$. When within a block this is thus the general formula.  It now follows that
\begin{eqnarray}
\langle \psi_m | \hat{\Delta}_C | \psi_m\rangle & = & \sum_{a,  b}  (\sigma_a^m)^* \sigma_b^m  \langle m;a | \Delta |m; b\rangle   \nonumber \\
& = & \sum_{k=1}^{N-m} \sum_{a,b} k (\sigma_a^m)^{*} \sigma_b^m ( \alpha^{m+k}_{m;a} )^* \alpha^{m+k}_{m; b}    \nonumber \\
& = & \sum_{k=1}^{N-m}  k |s^{m; m+k}|^2    \nonumber 
\end{eqnarray}
where we have defined 
$$ s^{m; m+k}  =  \sum_{a} \sigma_a^m \alpha^{m+k}_{m;a}    $$
which is again a sum of strictly non-negative terms. \\ \\
The general eigenstate can actually be recast as 
$$ |\psi_m\rangle = \sum_{a}  \sigma_a^m \sum_{q = m}^N  \alpha_{m;a}^{m+k} |K_q \rangle  = \sum_{k=0 }^{N-m} s^{m ; m+k} |K_{m+k}\rangle $$
so that the result of this subsection is essentially the same as that of the previous subsection.  The lesson is that it is only the eigenvalues of $\hat{C}_1 + \hat{C}_2$, the eigenvalues of $\hat{C}_{12}$ and how the corresponding eigenvectors are coupled that are ultimately important.  

\subsubsection{Superposition of eigenstates}

We now turn our attention to a superposition of eigenstates of $\hat{C}_1 + \hat{C}_2$, the most general state that we can consider.  We will start with the simplest example namely target state that is a superposition of two eigenstates of $\hat{C}_1 + \hat{C}_2$
\begin{equation}
|\psi_{mq}\rangle = c_m |\psi_m\rangle + c_q |\psi_q\rangle 
\end{equation}
Without loss of generality we will assume that $q > m$.  The last matrix elements we need are
\begin{equation}
\langle \psi_m | \hat{\Delta}_C | \psi_q\rangle = \sum_{k=1}^{N-q}  k \left(s^{m; q+k}\right)^* \left( s^{q; q+k}  \right)   \nonumber 
\end{equation}
Note that the index corresponding to the Krylov basis vectors must be at least $q$.  As a reminder, we have that
\begin{eqnarray}
\langle \psi_q | \hat{\Delta}_C | \psi_q\rangle = \sum_{k=1}^{N-q}  k \left| s^{q; q+k}  \right|^2   \nonumber
\end{eqnarray}
but for $\langle \psi_m | \hat{\Delta}_C | \psi_m\rangle$ we have
\begin{eqnarray}
\langle \psi_m | \hat{\Delta}_C | \psi_m\rangle & = & \sum_{k=1}^{N-m}  k \left| s^{m; m+k}\right|^2    \nonumber \\
& = & \sum_{k=1}^{q-m}  k \left| s^{m; m+k} \right |^2 +  \sum_{l=1}^{N-q}  (l + q-m) \left|  s^{m; q+l} \right |^2   \nonumber \\ 
& = & \sum_{l=1}^{N-q}  l \left|  s^{m; q+l} \right |^2 +  \sum_{k=1}^{q-m}  k \left| s^{m; m+k} \right |^2  + (q-m)\sum_{l=1}^{N-q} \left|  s^{m; q+l} \right |^2    \nonumber \\
& \geq & \sum_{k=1}^{N-q}  k \left|  s^{m; q+k} \right |^2    \nonumber
\end{eqnarray}
In other words, since $m < q$ the overlap $\langle \psi_m | \hat{\Delta}_C | \psi_m\rangle$ involves additional (positive) terms that involve Krylov basis vectors with index smaller than $q$.  It now follows, for any values $c_m,c_q$, that
\begin{eqnarray}
& & \langle \psi_{mq} | \hat{\Delta}_C | \psi_{mq}   \rangle   \nonumber \\
& = & |c_m|^2 \langle \psi_m | \hat{\Delta}_C |\psi_m\rangle + c_m (c_q)^* \langle \psi_q | \hat{\Delta}_C |\psi_m\rangle  + (c_m)^* c_q \langle \psi_m | \hat{\Delta}_C |\psi_q\rangle + |c_q|^2 \langle \psi_q | \hat{\Delta}_C |\psi_q\rangle    \nonumber \\
& \geq & \sum_{k=1}^{N-q} k |c_m s^{m; q+k} + c_q s^{q; q+k}    |^2     \nonumber
\end{eqnarray}
Note that the above demonstrates that the excess complexity is at least its contribution restricted to the Krylov states with $n > q$.  This is a general pattern that allows for proof by iteration.  For example, consider the next level, a state that is a superposition of three eigenstates 
$$|\psi_{m q r}\rangle =  c_m |\psi_m\rangle + c_q |\psi_q\rangle + c_r |\psi_r\rangle$$
with $r > q > m$.  We already know that for any values of $c_m, c_q$ the terms coming from the first two states above will give a non-negative value, more specifically a sum of terms.  To prove that $\langle \psi_{mqr} |\hat{\Delta}_C |   \psi_{mqr} \rangle$ is not negative we will not need all of the terms in the sum since  
$$ \langle \psi_{mq} |\hat{\Delta}_C | \psi_r\rangle = \sum_{k=1}^{N-r} k (c_m s^{m; r+k} + c_q s^{q; r+k}    )^* s^{r; r+k}    $$
and $\langle \psi_{r} | \hat{\Delta}_C | \psi_{r}   \rangle$ involve overlaps of Krylov basis vectors starting from $r$.  After dropping the extra (postive) terms from $\langle \psi_{mq} |\hat{\Delta}_C |   \psi_{mq} \rangle$ it follows that
$$\langle \psi_{mqr} |\hat{\Delta}_C |   \psi_{mqr} \rangle  \geq \sum_{k=1}^{N-r} k |c_m s^{m; r+k} + c_q s^{q; r+k} + c_r s^{r; r+k}|^2$$
We can now proceed to add another eigenvector to our superposition with eigenvalue greater than $r$ and so on.  By iteration we thus prove that
\begin{equation}
\langle \Psi | \hat{C}_{12} - \hat{C}_1 - \hat{C}_2|\Psi\rangle \geq 0
\end{equation}
for any state $|\Psi\rangle$ supported on the Krylov basis.  A general tensor product of Hamiltonians may be combined piece-by-piece to generalise this inequality.  For example
$$\langle \Psi | \hat{C}_{123}|\Psi\rangle \geq \langle \Psi | \hat{C}_{12} + \hat{C}_3|\Psi\rangle   \geq  \langle \Psi | \hat{C}_{1} + \hat{C}_2 + \hat{C}_3|\Psi\rangle   $$
By another iterative argument we can prove the general inequality that 
$$\langle \Psi | \hat{C}_{123 \cdots n} - \sum_{i=1}^n \hat{C}_i | \Psi\rangle \geq 0 $$
In what follows we will mainly restrict to the tensor product of two subsystems, but the statements generalize easily to a larger number of subsystems.  

\subsection{Matrix elements in the Krylov basis}

Expressed in the Krylov basis the matrix elements of $\hat{\Delta}_C$ provide some additional insight.  Here we will cast the Krylov basis vectors as 
$$ |K_n\rangle = \sum_{  a = 0  }^n     \rho_{n, n-a}   \sum_{ k = 0}^{n-a}   c_{n; a; k} |n-a; k\rangle $$
normalised such that $ \sum_{a=0}^n (\rho_{n,n-a})^2 = 1  $ and $\sum_{k=0}^{n-a} |c_{n; a; k}|^2 = 1  $.  The $(\rho_{n, a})^2$ is thus the probability amplitude squared for an eigenstate of $\hat{C}_{12}$ with eigenvalue $n$ to be contained in an eigenstate of $\hat{C}_1 + \hat{C}_2$.  We have that
\begin{eqnarray}
\langle K_n | \hat{\Delta}_C | K_n\rangle  & = & n - \sum_{ a=0 }^n (\rho_{n; n-a})^2 (n-a) \nonumber \\
& = & \sum_{ a=0 }^n (\rho_{n; n-a})^2 a    \nonumber \\
& = & \sum_{b=0}^{min(n, N_1+N_2)} (n-b) (  \rho_{n;b}    )^2  
\end{eqnarray}
The diagonal elements of the excess complexity operator thus measure the mean deviation of the Krylov basis index from the sum of the tensor product indices.  To quantify the off-diagonal elements we note that 
\begin{eqnarray}
& & \sum_{m}     \langle K_n| \hat{C}_{12} - \hat{C}_{1} - \hat{C}_2 |K_m\rangle  \langle K_m |\hat{C}_{12} - \hat{C}_{1} - \hat{C}_2 |K_n\rangle    \nonumber \\
& = &  \langle K_n| (\hat{C}_{12} - \hat{C}_{1} - \hat{C}_2)^2 |K_n\rangle    \nonumber \\
& = & \sum_{a=0}^n  \left(  n^2 (\rho_{n;n-a})^2 - 2 n(n-a) (\rho_{n;n-a})^2 + (n-a)^2  (\rho_{n;n-a})^2  \right)    \nonumber \\
& = & \sum_{a=0}^n a^2 (\rho_{n; n-a})^2 = \sum_{b=0}^{min(n, N_1+N_2)} (n-b)^2 (\rho_{n;b})^2 \end{eqnarray}
It now follows that
\begin{equation}
\sum_{m\neq n}    |\langle K_n | \hat{C}_{12} - \hat{C}_1 - \hat{C}_2 | K_n\rangle |^2 = \frac{1}{2}\sum_{k,l} (k-l)^2 (\rho_{n;k})^2 (\rho_{n;l})^2
\end{equation}
The sum of the magnitude squared of the off-diagonal matrix elements in each row or column is thus the variance of the difference between the Krylov basis and tensor product indices.  

\subsection{Time-evolved reference state and synchrony}

The excess complexity operator is positive definite and thus a positive quantity for \textbf{any} state supported on the Krylov basis.  A case of particular interest is to consider the target state as the time-evolved reference state.  This state may be expressed in two different ways
\begin{eqnarray}
e^{-i t (H_1 + H_2)} |\phi_1\rangle \otimes |\phi_2\rangle & = & \sum_{m,m' } \phi_{1,m}(t)  \phi_{2,m'}(t) |K_{1,m}\rangle \otimes |K_{2,m'}\rangle     \nonumber \\
& = & \sum_{n} \Psi(t) |K_n\rangle     \nonumber   
\end{eqnarray}
using the Krylov bases for the two subsystems and the total system respectively.  The excess complexity of the time-evolved reference state is zero if we have that
$$ |K_{n}\rangle = \sum_{k=0}^n c_{n-k, k}  |K_{1,n-k} \rangle \otimes |K_{2,k}\rangle \ \ \ \Rightarrow \ \ \ C_{12}(t) = C_{1}(t) + C_2(t)      $$
which we will now demonstrate.  Assume that the $n'th$ Krylov basis vector is a linear combination of states with subscript adding to $n$.  This implies that 
\begin{eqnarray}
\sum_{n} \Psi_n(t) |K_n\rangle & = & \sum_{n} \sum_{k=0}^n \Psi_{n}(t) c_{n-k,k} |K_{1,n-k}\rangle \otimes |K_{2,k}\rangle      \nonumber \\
& = & \sum_{m,m'} \Psi_{m+m'}(t) c_{m,m'} |K_{1,m}\rangle \otimes |K_{2,m'}\rangle     \nonumber     \nonumber \\
\Rightarrow \Psi_{m+m}(t) c_{m,m'} & = & \psi_{1,m}(t) \psi_{2,m'}(t)    \nonumber
\end{eqnarray}
This identity is the statement of synchrony on the level of the time-evolved reference state and implies that the complexity of the combined system is 
\begin{eqnarray}
C_{12}(t) & = & \sum_{n} n |\Psi_{n}(t)|^2    \nonumber \\
& = & \sum_{n} \sum_{k} n |\Psi_n(t)|^2 |c_{n-k, k}|^2     \nonumber \\
& = & \sum_{m,m'} (m+m') |\Psi_{m+m'}(t)|^2 |c_{m,m'}|^2    \nonumber \\
& = & \sum_{m,m'} m |\psi_{1,m}|^2 |\psi_{2,m'}|^2 + \sum_{m,m'} m' |\psi_{1,m}|^2 |\psi_{2,m'}|^2 \nonumber \\
& = & \sum_{m} m |\psi_{1,m}|^2 + \sum_{m'} m' |\psi_{2,m'}|^2    \nonumber \\
& = & C_{1}(t) + C_{2}(t)
\end{eqnarray}
as it should.  For a general state built from the Krylov basis we have that the inequality is saturated if and only if
$$ |K_n\rangle =  \sum_{k=0}^n c_{n-k,k} |K_{1,n-k}\rangle \otimes |K_{2,k}\rangle \ \ \ \Leftrightarrow \ \ \   \langle K_n|  \hat{C}_{1}  + \hat{C}_2 | K_n\rangle =  n  $$
This is in line with our results in the previous subsections.  For a general state, the above statement is that of synchrony.  \\ \\
The cleanest way to define synchrony, however, is to make use of the decomposition of the Hamiltonian that results from the Lanczos algorithm.  For the two subsystems we have the decomposition
$$ H_{j} = L_{j,+} + L_{j,0} + L_{j,-} $$
representing the entries above the diagonal, on the diagonal below the diagonal respectively when written in matrix form in their respective Krylov bases.  Similarly, the total system Hamiltonian permits a decomposition of this form.  We thus have two (potentially competing) decompositions of the total system Hamiltonian 
$$ H = L_{-} + L_0 + L_{+} = ( L_{1,-} + L_{2,-}) + (L_{1,0} + L_{2,0}) + ( L_{1,+} + L_{2,+})   $$
 The two systems evolve synchronously if these decompositions are in harmony i.e. if one has that\footnote{This statement implies that $L_{-}$ must also satisfy a similar identity and thus that $L_0$ must also satisfy a similar identity.}
\begin{eqnarray}
L_{+} & = & L_{1,+} + L_{2,+}    \label{synchroCondition}
\end{eqnarray}
This leads directly to the identification
$$ |K_n\rangle =  \frac{(L_{1,+} + L_{2,+})^n |K_0\rangle }{ \sqrt{ \langle K_0 | (L_{1,-} + L_{2,-})^n  (L_{1,+} + L_{2,+})^n |K_0\rangle    } }   $$
and gives zero excess complexity.

\subsection{Some comments on the Krylov basis}

It is clear that the excess complexity is due to the Krylov basis vectors deviating from the pattern that the $n$'th basis vector is a linear superposition of the subsystem basis vectors with index summing to $n$. This first few Krylov basis vectors provide some insight as to how this happens.  We will take the Lanczos coefficients associated to $H_j$ to be $a_{j,n}$ and $b_{j,n}$.  By virtue of the Lanczos algorithm we know that
\begin{eqnarray}
|K_0\rangle & = & |K_{1,0}\rangle \otimes |K_{2,0}\rangle    \nonumber \\
|K_1\rangle & = & \frac{b_{1,1}}{\sqrt{ (b_{1,1})^2 + (b_{2,1})^2   }} |K_{1,1}\rangle \otimes |K_{2,0}\rangle +  \frac{b_{2,1}}{\sqrt{ (b_{1,1})^2 + (b_{2,1})^2   }}  |K_{1,0}\rangle \otimes |K_{2,1}\rangle     \nonumber 
\end{eqnarray}
It thus follows that the first vector that can be responsible for a non-zero excess complexity is the vector orthogonal to $|K_1\rangle$
\begin{equation}
\frac{b_{2,1}}{\sqrt{ (b_{1,1})^2 + (b_{2,1})^2   }} |K_{1,1}\rangle \otimes |K_{2,0}\rangle -  \frac{b_{1,1}}{\sqrt{ (b_{1,1})^2 + (b_{2,1})^2   }}  |K_{1,0}\rangle \otimes |K_{2,1}\rangle     \nonumber
\end{equation}
This state first starts appearing in $|K_2\rangle$ if $a_{1,1} \neq a_{2,1}$ and in $|K_3\rangle$ in most other instances.  A consequence of this is that the excess complexity of the time-evolved reference state $e^{-iHt}\ket{\phi_r}$
\begin{equation}
\Delta C(t) = C_{12} - (C_1 + C_2)
\end{equation}
grows either as  $\Delta C(t) \sim t^4$ or $\Delta C(t) \sim t^6$ at leading order, so that at early times the two complexities are nearly additive. At late times, however, the interference between distinct Krylov sectors leads to a bounded, oscillatory excess depending on the coupling coefficients (in the case of finite-dimensional Krylov spaces). Thus, the excess complexity is \emph{bounded} and oscillatory for finite-dimensional systems, reflecting coherent interference between independent Krylov chains.

\section{An $su(2)$ example}

Our discussion thus far has been completely general.  We will now focus on an illustrative example of what is described above which is the tensor product of two Hamiltonians that are elements of the $su(2)$ algebra.  Specifically we will be interested in the Hamiltonian 
$$H = \alpha_1(  J_{1,+} + J_{1,-}   ) + \alpha_2(J_{2,+} + J_{2,-})$$
where the operators above satisfy the usual $su(2)$ commutation relations
\begin{eqnarray}
\left[ J_{i,+}, J_{j,-}   \right] & = & 2 \delta_{i,j} J_{i,0}    \nonumber \\
\left[ J_{i,0}, J_{j, \pm}   \right] & = & \pm \delta_{ij} J_{i,\pm}      \nonumber
\end{eqnarray}
As reference state we will choose 
$$ |\phi_r\rangle = | j_{1}, -j_1; j_{2}, -j_2\rangle$$
which is the lowest weight state that satisfies
$$ J_{i,-}|\phi_r\rangle = 0.$$
The Krylov basis generated by the action of $H_{i}$ on the reference state is known \cite{Caputa:2021sib, Chattopadhyay:2023fob} to be 
\begin{equation}
|K_{j,n}\rangle = (J_{j,+})^n |\phi_r\rangle
\end{equation}
up no normalisation and the Krylov complexity operator is
$$ \hat{C}_i = J_{i, 0} + j_i   $$
For the $su(2)$ example it will thus be possible to understand the excess complexity in terms the failure of the $n$'th Krylov basis vector to be an eigenstate of $\sum_{i} (J_{i,0} + j_{i,0})$ with eigenvalue $n$.

\subsection{$j_1 = j_2 = \frac{1}{2}$}

\begin{figure}[h!]
    \centering
    \includegraphics[width=0.6\linewidth]{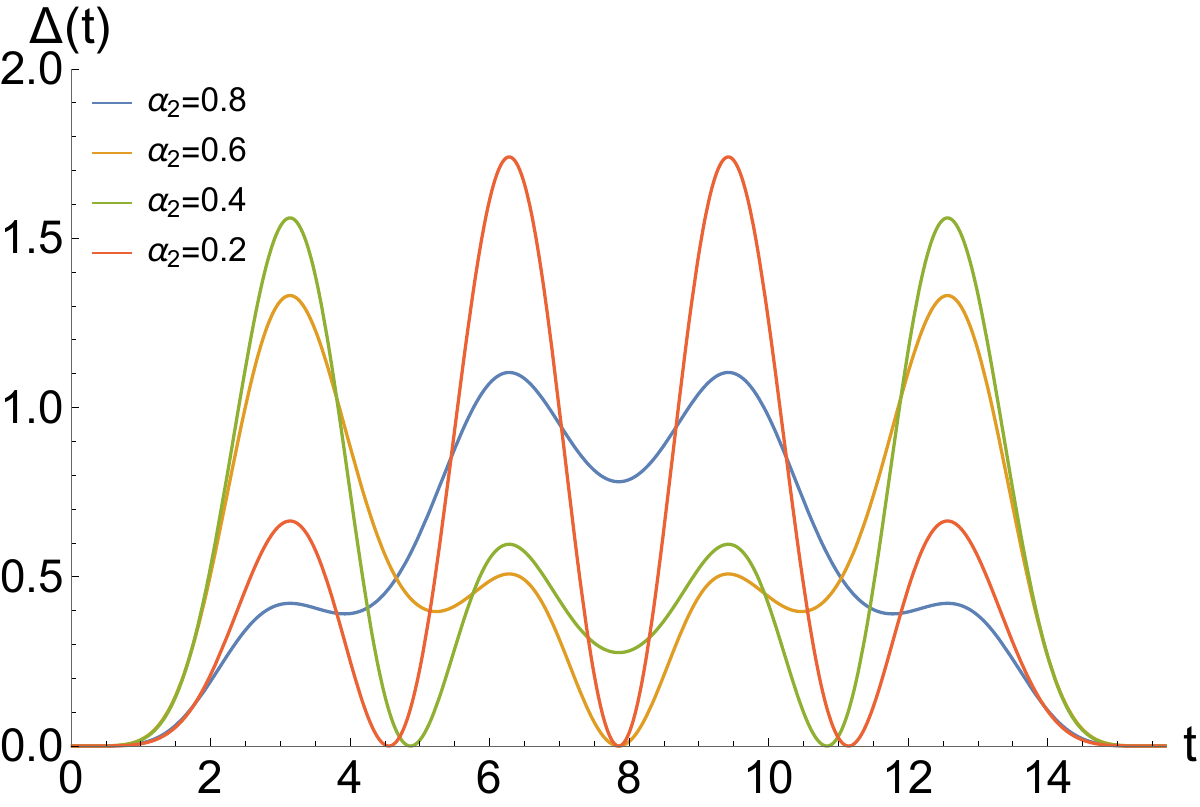}
    \caption{The excess complexity of the time-evolved lowest weight state for $j_1 = j_2 = \frac{1}{2}$, $\alpha_1 = 1$ and various values of $\alpha_2$.  The excess complexity is always positive, bounded and oscillatory.}
    \label{fig:jhalfhalfPlot}
\end{figure}

The Krylov basis for the combined Hamiltonian (for $j_1 = j_2 = \frac{1}{2}$) is given by 
\begin{eqnarray}
|K_0\rangle & = & |\phi_r\rangle     \nonumber \\
|K_1) & = & (\alpha_1 J_{1,+} + \alpha_2 J_{2,+})|\phi_r\rangle     \nonumber \\
|K_2) & = & (\alpha_1 J_{1,+} + \alpha_2 J_{2,+})^2|\phi_r\rangle  = 2 \alpha_1 \alpha_2 J_{1,+} J_{2,+}|\phi_r\rangle    \nonumber \\
|K_3) & = & (\alpha_2 J_{1,+} - \alpha_1 J_{2,+})|\phi_r\rangle    \nonumber
\end{eqnarray}
We find the following complexity operators for the different tensor product indices
\begin{eqnarray}
\hat{C}_1 & = & |K_{1,1}\rangle \otimes |K_{2,0}\rangle \langle K_{1,1}| \otimes \langle K_{2,1}| +  |K_{1,1}\rangle \otimes |K_{2,1}\rangle \langle K_{1,1}| \otimes \langle K_{2,1}|    \nonumber \\
\hat{C}_2 & = & |K_{1,0}\rangle \otimes |K_{2,1}\rangle \langle K_{1,0}| \otimes \langle K_{2,1}| +  |K_{1,1}\rangle \otimes |K_{2,1}\rangle \langle K_{1,1}| \otimes \langle K_{2,1}|      \nonumber
\end{eqnarray}
so that
\begin{eqnarray}
\hat{C}_1 + \hat{C}_2 & = & |K_{1,1}\rangle \otimes |K_{2,0}\rangle \langle K_{1,1}| \otimes \langle K_{2,1}| + |K_{1,0}\rangle \otimes |K_{2,1}\rangle \langle K_{1,0}| \otimes \langle K_{2,1}|   \nonumber \\
& + & 2|K_{1,1}\rangle \otimes |K_{2,1}\rangle \langle K_{1,1}| \otimes \langle K_{2,1}|    \nonumber
\end{eqnarray}
This may be recast as 
\begin{eqnarray}
\hat{C}_1 + \hat{C}_2 & = & |K_1\rangle\langle K_1| +  2|K_2\rangle\langle K_2| +  |K_3\rangle\langle K_3|
\end{eqnarray}
while 
\begin{equation}
\hat{C}_{12} =  |K_1\rangle\langle K_1| +  2|K_2\rangle\langle K_2| +  3|K_3\rangle\langle K_3|.
\end{equation}
This makes it clear that for \textbf{any} choice of target state, one has that 
\begin{equation}
\langle \phi_t| \hat{C}_{12} |\phi_t\rangle \geq \langle \phi_t| \hat{C}_{1} |\phi_t\rangle + \langle \phi_t| \hat{C}_{2} |\phi_t\rangle
\end{equation}
It is also clear that the excess complexity comes about due to the overlap of the target state with the Krylov basis vector $|K_3\rangle$.  \\ \\
An important example target state is that of the time-evolved reference state, $|t\rangle = e^{- i t H}|\phi_r\rangle$.  For this state we can compute
\begin{equation}
\langle t |\hat{C}_{12} - (\hat{C}_1 + \hat{C}_2)|t\rangle = 2\frac{(\alpha_2 \cos(\alpha_2 t) \sin(\alpha_1 t) - \alpha_1 \cos(\alpha_1 t) \sin(\alpha_2 t)  )^2     }{\alpha_1^2 + \alpha_2^2}
\end{equation}
As a series expansion in $t$, the first non-vanishing contribution to the above is at order $t^6$ so that the two quantities is closely matched at early times.  Note that, if $|\alpha_1| = |\alpha_2|$ or either $\alpha_1 = 0$ or $\alpha_2 = 0$ then the excess complexity is zero.  We have plotted the excess complexity for this case in Fig. (\ref{fig:jhalfhalfPlot})

\subsection{$j_1 = 1  \ \ ; \ \  j_2 = \frac{1}{2}  $}

\begin{figure}
    \centering
    \includegraphics[width=0.6\linewidth]{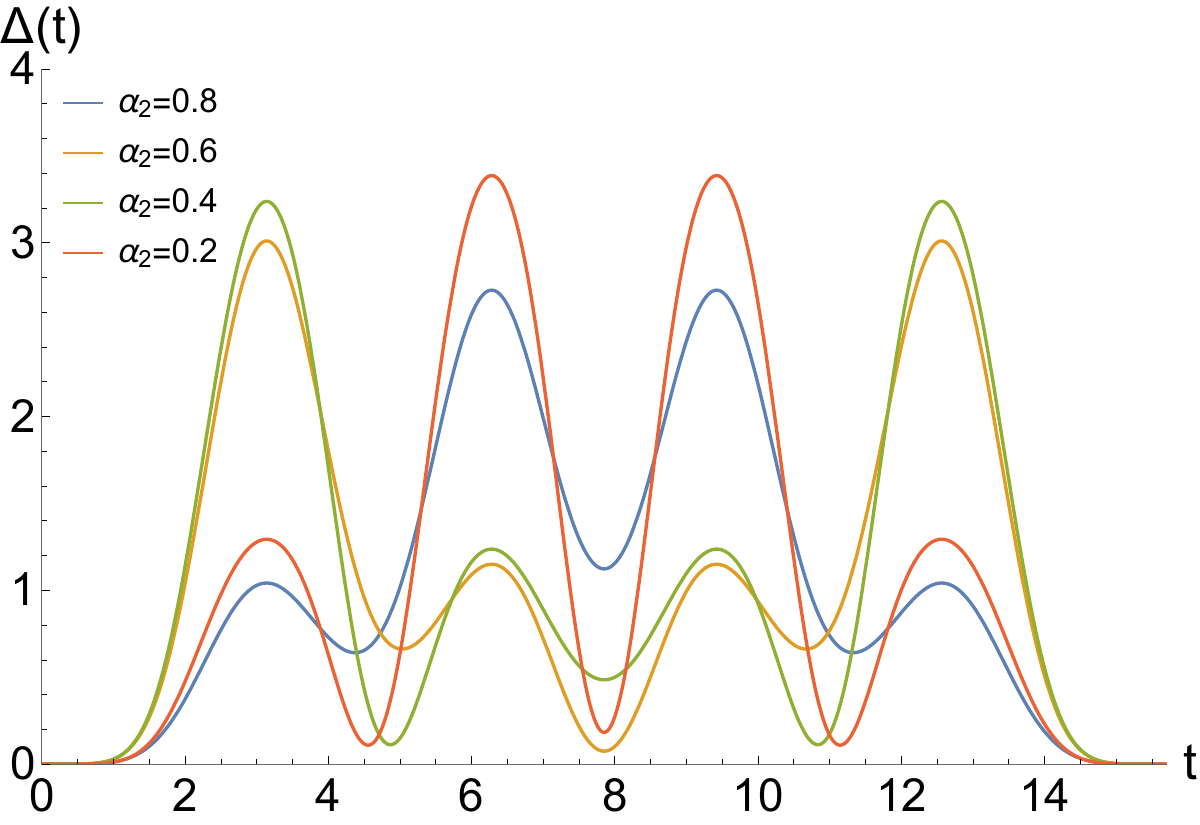}
    \caption{The excess complexity of the time-evolved lowest weight state for $j_1 = 1; j_2 = \frac{1}{2}$, $\alpha_1 = 1$ and various values of $\alpha_2$.  The excess complexity is always positive, bounded and oscillatory.}
    \label{fig:jonejhalf}
\end{figure}

The first two Krylov basis vectors for $j_1 = 1$ and $j_2 = \frac{1}{2}$ is given by 
\begin{eqnarray}
|K_0\rangle & = & |\phi_r\rangle     \nonumber \\
|K_1\rangle & = & (\alpha_1 J_{1,+}  +  \alpha_2 J_{2,+}  )|\phi_r\rangle     \nonumber \\
|K_2) & = & (\alpha_1 J_{1,+}  +  \alpha_2 J_{2,+}  )^2|\phi_r\rangle  
\end{eqnarray}
Indeed, for our choice of Hamiltonian this is the first three basis vectors for general $j_1, j_2$.  The next Krylov basis vector is given by the superposition
\begin{eqnarray}
|K_3) =  (\alpha_1 J_{1,+}  +  \alpha_2 J_{2,+}  )^3|\phi_r\rangle + \frac{2 \alpha_1 \alpha_2 (\alpha_2^2 - \alpha_1^2)}{2 \alpha_1^2 + \alpha_2^2}    (\alpha_2 J_{1,+}  -  2 \alpha_2 J_{2,+}  )|\phi_r\rangle    \nonumber
\end{eqnarray}
The second vector above is responsible for excess complexity of the tensor product.  It is the state with the same $J_0$ eigenvalue as $|K_1\rangle$ and orthogonal to it.  The $n$'th Krylov basis vector will always contain a component $(\alpha_1 J_{1,+}  +  \alpha_2 J_{2,+}  )^n|\phi_r\rangle$ for $n \leq 2 j_1 + 2 j_2$.  Beyond this value for $n$ further Krylov basis vectors are still generated by the Hamiltonian.  
\\ \\
For this value of $j_1, j_2$ the last two Krylov basis vectors are 
$$ |K_4) = (\alpha_1 \alpha_2 J_{1,+} + (\alpha_1^2 + \alpha_2^2)J_{2,+})( \alpha_1 J_{1,+} + \alpha_2 J_{2,+}  ) |\phi_r\rangle  $$
which is the state with the same $J_0$ eigenvalue as $|K_2)$ orthogonal to it and, finally,
$$ |K_5) =  (\alpha_1 J_{1,+}  +  \alpha_2 J_{2,+}  )^3|\phi_r\rangle - \frac{6\alpha_1^3 \alpha_2 (4 \alpha_1^2 - 5 \alpha_1^2 \alpha_2^2 + \alpha_2^4)}{5(20 \alpha_1^2 + 5 \alpha_1^2 \alpha_2^2 + 2\alpha_2^4)}  (\alpha_2 J_{1,+}  -  2 \alpha_2 J_{2,+}  )|\phi_r\rangle $$
which is the same superposition as $|K_3)$ but orthogonal to it.  Both of these vectors contribute towards excess complexity.   \\ \\
To quantify the excess complexity we may either consider the expectation value of $J_z + j_1 + j_2 = \hat{C}_1 + \hat{C}_2$ w.r.t. the Krylov basis vectors or compute the complexity of the individual tensor product state.  The former yields 
\begin{eqnarray}
\langle K_0 | \hat{C}_1 + \hat{C}_2 | K_0\rangle & = & 0   \nonumber \\
\langle K_1 | \hat{C}_1 + \hat{C}_2 | K_1\rangle & = & 1   \nonumber \\
\langle K_2 | \hat{C}_1 + \hat{C}_2 | K_2\rangle & = & 2   \nonumber \\
\langle K_3 | \hat{C}_1 + \hat{C}_2 | K_3\rangle & = & 3 - \frac{4 (\alpha_1^2 - \alpha_2^2)^2}{20 \alpha_1^2 + 5 \alpha_1^2 \alpha_2^2 + 2 \alpha_2^4}     \nonumber \\
\langle K_4 | \hat{C}_1 + \hat{C}_2 | K_4\rangle & = & 2   \nonumber \\
\langle K_5 | \hat{C}_1 + \hat{C}_2 | K_5\rangle & = & 1 + \frac{4 (\alpha_1^2 - \alpha_2^2)^2}{20 \alpha_1^2 + 5 \alpha_1^2 \alpha_2^2 + 2 \alpha_2^4} \nonumber 
\end{eqnarray}
for the diagonal elements and two equal off-diagonal elements
$$ \langle K_3 | \hat{C}_1 + \hat{C}_2 | K_5\rangle = \langle K_5 | \hat{C}_1 + \hat{C}_2 | K_3\rangle = \frac{6 \sqrt{4 \alpha_1^2 + 2 \alpha_2^2} \alpha_1 (\alpha_2^2 - \alpha_1^2 )}{20 \alpha_1^4 + 5 \alpha_1^2 \alpha_2^2 + 2 \alpha_4^2}  $$
For the latter computation it is useful to organise the states according to the $J_z + j_1 + j_2$ eigenvalue.  The reference state is trivial.  For eigenvalue $1$ we have two states and the most and least complex combinations are 
\begin{eqnarray}
\frac{\langle \phi_r | ( \alpha_1 J_{1,-} + \alpha_2 J_{2,-}) \hat{C}_{12} ( \alpha_1 J_{1,+} + \alpha_2 J_{2,+})| \phi_r  \rangle     }{ \langle \phi_r | ( \alpha_1 J_{1,-} + \alpha_2 J_{2,-}) ( \alpha_1 J_{1,+} + \alpha_2 J_{2,+})| \phi_r  \rangle      }  & = & 1    \nonumber \\
\frac{\langle \phi_r | ( \alpha_2 J_{1,-} - 2 \alpha_1 J_{2,-}) \hat{C}_{12} ( \alpha_2 J_{1,+} - 2 \alpha_1 J_{2,+})| \phi_r  \rangle     }{ \langle \phi_r | ( \alpha_2 J_{1,-} - 2 \alpha_1 J_{2,-})  ( \alpha_2 J_{1,+} - 2 \alpha_1 J_{2,+})| \phi_r  \rangle      }  & = & 5 - \frac{4 (\alpha_1^2 - \alpha_2^2)^2}{20 \alpha_1^4 + 5 \alpha_1^2 \alpha_2^2 + 2 \alpha_2^4}.    \nonumber
\end{eqnarray}
For eigenvalue $2$ the minimum complexity is $2$ and the most complex state in this subspace has complexity 4.  Finally, the highest weight state has complexity
$$  \langle j_1, j_1 ; j_2, j_2 |    \hat{C}_{12} | j_1, j_1 ; j_2, j_2\rangle = 3 +   \frac{4 (\alpha_1^2 - \alpha_2^2)^2}{20 \alpha_1^4 + 5 \alpha_1^2 \alpha_2^2 + 2 \alpha_2^4 }. $$
These results are completely in line with our expectations.  The Hilbert space consists of six states with $J_z + j_1 + j_2$ eigenvalues of $0, 1, 1, 2, 2, 3$ respectively.  The Krylov basis (for general values of $\alpha_1, \alpha_2$) explores all these states with deviations expected from $|K_3\rangle$ onwards.  The trace of this operator expressed in the Krylov basis should equal the sum of the eigenvalues, so that these deviations should cancel. \\ \\
The excess complexity operator, $\hat{\Delta}_C$,  has three eigenvectors with eigenvalue $0$ namely $|K_0\rangle, |K_1\rangle$ and $|K_2\rangle$.  The state $|K_4\rangle$ is an eigenvector with eigenvalue $2$ and, finally, it has two other eigenvalues. The excess complexity operator, $\hat{\Delta}_C$,  has three eigenvectors with eigenvalue $0$ namely $|K_0\rangle, |K_1\rangle$ and $|K_2\rangle$.  The state $|K_4\rangle$ is an eigenvector with eigenvalue $2$ and, finally, it has two other eigenvalues that satisfy
\begin{eqnarray}
\lambda_1 + \lambda_2 & = & 4    \nonumber \\
\lambda_1 \lambda_2 & = & \frac{8(\alpha_1^2 - \alpha_2^2)^2}{20 \alpha_1^4 + 5 \alpha_1^2 \alpha_2^2 + 2\alpha_2^4}     \nonumber
\end{eqnarray}
and are thus positive.  We have plotted the excess complexity for this case in Fig. (\ref{fig:jonejhalf}).

\subsection{Near-synchronous evolution}

\begin{figure}
    \centering
    \includegraphics[width=0.6\linewidth]{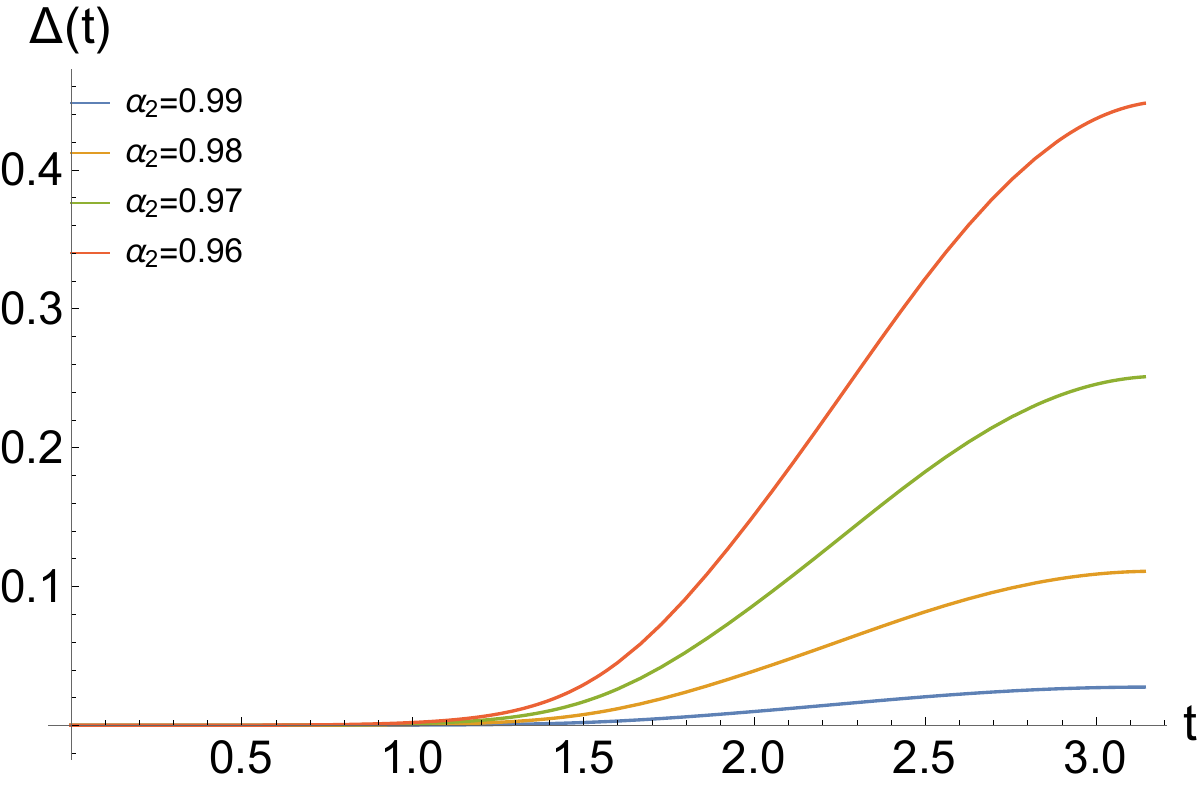}
    \caption{The excess complexity for $j_1 = j_2 = 2$,  $\alpha_1 = 1$ and various values of $\alpha_2 \approx 1$.  At small times the excess complexity remains small relative to the complexity (which has a maximal value of approximately $2(j_1 + j_2)$)}
    \label{fig:earlyTimes}
\end{figure}

\begin{figure}
    \centering
    \includegraphics[width=0.6\linewidth]{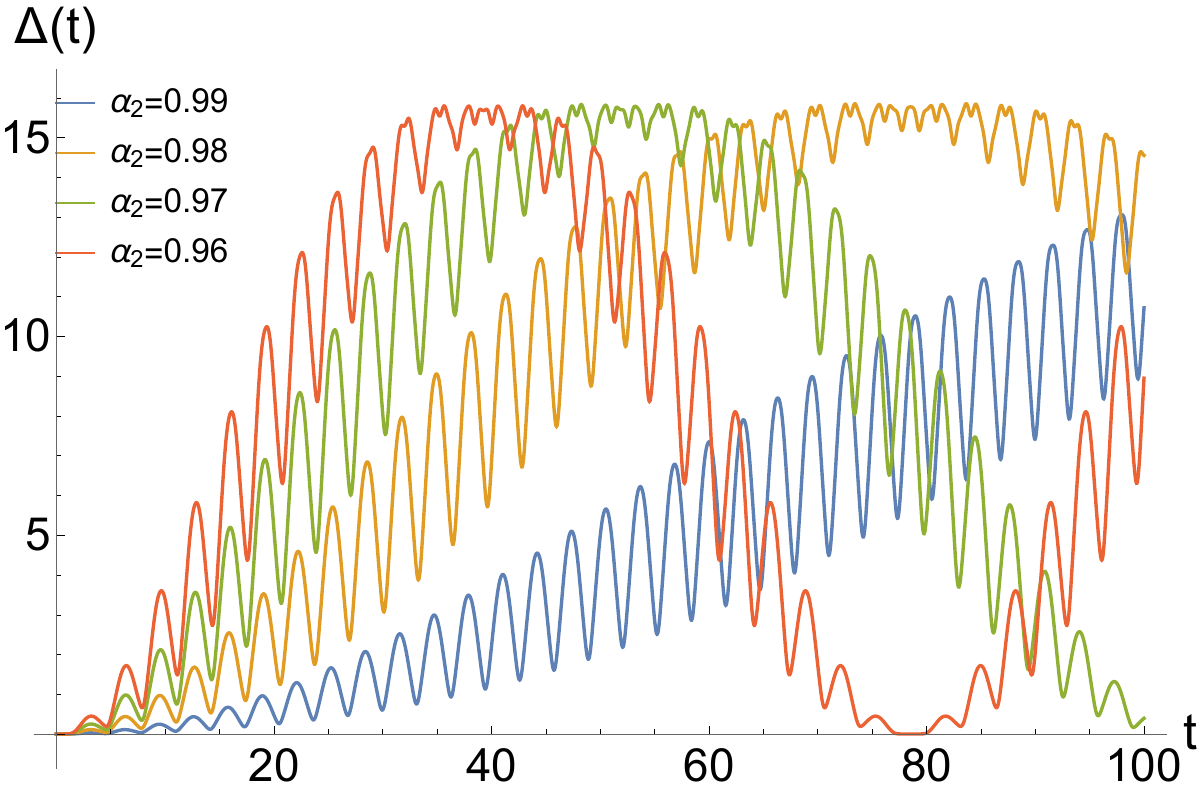}
    \caption{The excess complexity for $j_1 = j_2 = 2$,  $\alpha_1 = 1$ and various values of $\alpha_2 \approx 1$.  At larger times the excess complexity becomes large for all cases away from the synchronous point, though the time-scale is longer for cases with evolution closer to synchrony.  }
    \label{fig:lateTimes}
\end{figure}

The excess complexity vanishes when $|\alpha_1| = |\alpha_2|$ in which case the subsystems evolve in synchrony and (\ref{synchroCondition}) is satisfied.  A natural question is what happens when the dynamics are near-synchronous i.e. $|\alpha_1| \approx |\alpha_2|$.   When discussing general $j_1, j_2$, the following identity proves useful
$$ e^{- i t (J_{+} + J_{-}) } |j,-j\rangle = \left( \frac{\sin(2 \alpha t)}{2 \alpha t}  \right)^j e^{ \log\left( \frac{ \tan(\alpha t)}{\alpha t}   \right) J_0  } e^{- i \alpha_1 t J_{+}} |j,-j\rangle   $$
Using this we are able to obtain
\begin{equation}
e^{- i t H}|\phi_r\rangle  =  N e^{\frac{1}{2}\log\left( \frac{\alpha_2 \tan(\alpha_1 t)}{\alpha 1 \tan(\alpha_2 t)} \right)(J_{1,0} - J_{2,0}) } e^{\sqrt{\frac{\tan(\alpha_1 t) \tan(\alpha_2 t) }{\alpha_1 \alpha_2} }( \alpha_1 J_{1,+} + \alpha_2 J_{2,+}   )  } |\phi_r\rangle
\end{equation}
The leading exponential above generates the states 
 $$|K_n^{0}\rangle = N ( \alpha_1 J_{1,+} + \alpha_2 J_{2,+}   )^n |\phi_r\rangle$$
which are eigenstates of $J_{1,0} + J_{2,0}$ with increasing eigenvalue as $n$ increases.  As we have already discussed, if the time-evolved state is supported on these states only, then the evolution is synchronous and we have zero excess complexity.  The second exponential does not change the eigenvalue of $J_{1,0} + J_{2,0}$, but mixes the time-evolved reference state within the eigenspaces themselves.  The evolution of the reference state is thus split into a part that advances into different eigenspaces of $J_{1,0} + J_{2,0}$ and another that explores these eigenspaces.  It is the latter that is responsible for generating excess complexity.   \\ \\
These states are mutually orthogonal and the matrix elements 
$$ \langle K_n^{0}| H |K_m^{0} \rangle$$
give rise to a tri-diagonal matrix.  Furthermore, the states $|K_n^{0}\rangle$ appear at order $H^n$ so that the states $|K_n^{0}\rangle$ have a complexity of at least $n$.  In the absence of the second exponential their complexity would be exactly $n$ and we would have no excess complexity. When the contributions from the exponentials are combined we are able to write
\begin{eqnarray}
 H^n |\phi_r\rangle  & = &   ( \alpha_1 J_{1,+} + \alpha_2 J_{2,+}   )^n|\phi_r\rangle    \nonumber \\
&  & + \sum_{m=0}^{n-2}\sum_{k=0}^{n-m} c_{n; m,k} (J_{1,0} - J_{2,0})^k ( \alpha_1 J_{1,+} + \alpha_2 J_{2,+}   )^m|\phi_r\rangle \nonumber
\end{eqnarray}
The first few Krylov basis vectors are then as follows
\begin{eqnarray}
|K_0\rangle & = & |\phi_r\rangle    \nonumber \\
|K_1) & = & ( \alpha_1 J_{1,+} + \alpha_2 J_{2,+}   )|\phi_r\rangle    \nonumber \\
|K_2) & = & ( \alpha_1 J_{1,+} + \alpha_2 J_{2,+}   )^2|\phi_r\rangle    \nonumber \\   
|K_3) & = & ( \alpha_1 J_{1,+} + \alpha_2 J_{2,+}   )^3|\phi_r\rangle  + \frac{2 \alpha_1 \alpha_2(\alpha_1^2 - \alpha_2^2)}{j_1 \alpha_1^2 + j_2 \alpha_2^2} ( \alpha_2 j_2 J_{1}^{+} - \alpha_1 j_1 J_{2}^{+}     )|\phi_r\rangle   \nonumber
\end{eqnarray}
Note that the deviation of the third Krylov basis vector from $|K^0_3\rangle$ is small if $|\alpha_1| \approx |\alpha_2|$ or $j_1, j_2$ are large.  The latter is in line with an argument put forward in \cite{Caputa:2022yju} where it was argued that, in the large $j$ limit, the complexities of different $su(2)$ factors are additive.  If the deviation is small, then $|K_n^0\rangle$ provides a suitable approximate Krylov basis \cite{Haque:2022ncl}.  \\ \\
When working at finite $j_1, j_2$ these considerations are valid for the first $2(j_1 + j_2) + 1$ Krylov basis vectors.  This first set of Krylov basis are approximately the simultaneous eigenstates of $J_{1,0} + J_{2,0}$ and the quadratic Casimir of the tensor product.  The next $2(j_1 + j_2) - 1$ Krylov basis vectors are again approximate eigenstes of both, but with a different eigenvalue for the Casimir.  The end result of these considerations is that, at early times, the excess complexity is small (compared to the total complexity).  However, at later times the time-evolved reference state may have a small support on the approximate Krylov basis vectors $|K_n^0\rangle$ and the excess complexity large.   We have plotted the early time and late time behavior for near-synchronous evolution cases in Figs. (\ref{fig:earlyTimes}) and (\ref{fig:lateTimes}).  

\section{Krylov Graphs and Their Interpretation}

\begin{figure}
\centering
\includegraphics[width=0.7\textwidth]{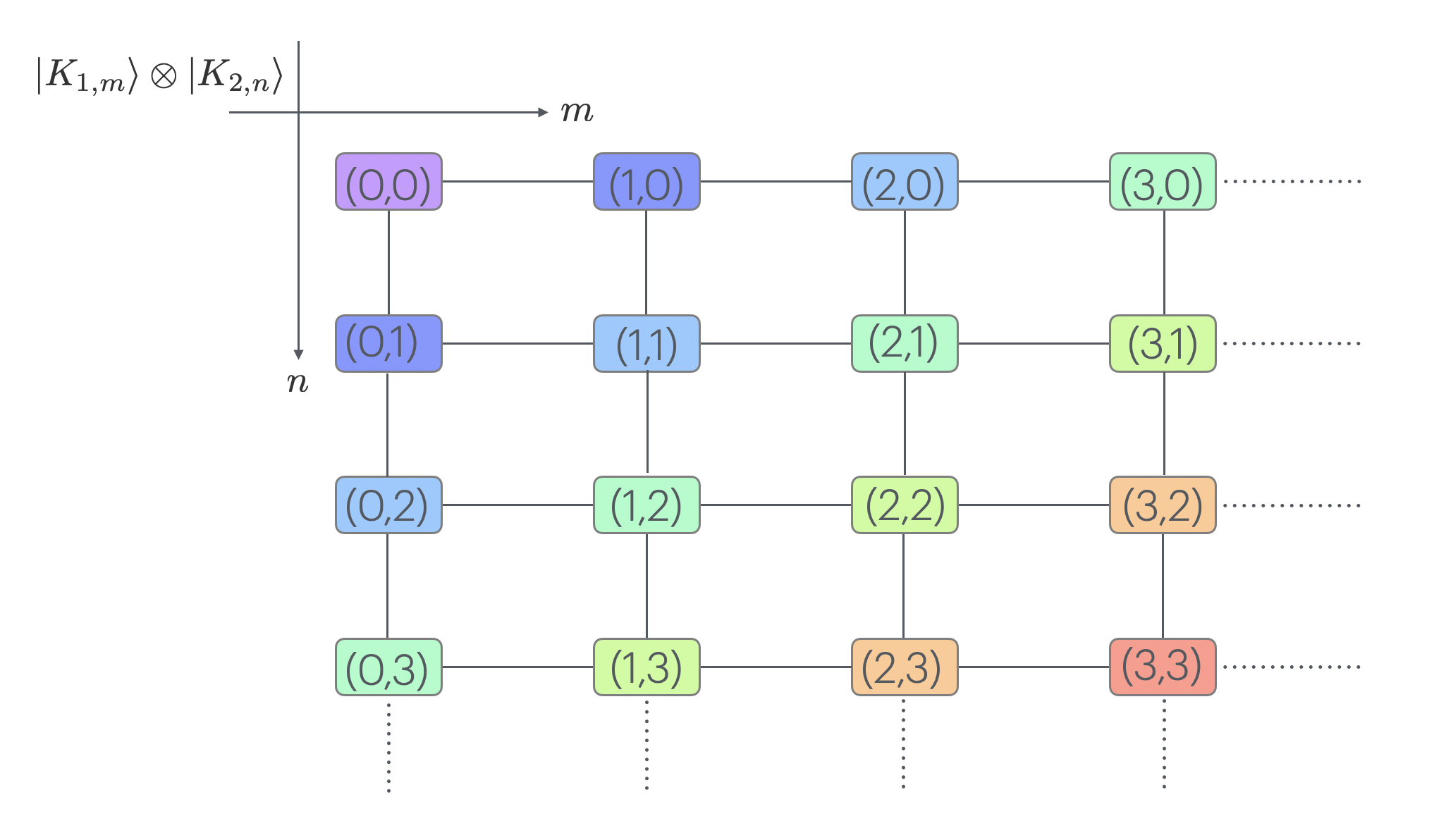}
\caption{The grid of the tensor product of Krylov vectors.  States with the same colour are eigenstates of $\hat{C}_1 + \hat{C_2}$ with the same eigenvalue. }
\label{KrylovGrid}
\end{figure}

From a computational perspective \cite{Beetar:2025mdz} the states $|K_n\rangle$ are obtained using two operations namely superposition and the combined system Hamiltonian.  The states $|m; k\rangle $ in contrast are obtained using three operations being superposition and both the subsystem Hamiltonians, $H_1$ and $H_2$.  The assignment of complexity is done accordingly for $\hat{C}_{12}$ and $C_{1} + \hat{C}_2$ by counting the total number of Hamiltonian applications.  Naturally, the complexity obtained with fewer operations is always higher for any target state.  We can organise the relevant states into a gird as in Fig. (\ref{KrylovGrid}).  As can be seen from \eqref{tensor-expansion}, the tensor product structure defines a two-dimensional Krylov graph--generalizing the 1-dimensional tight-binding Krylov chain--whose vertices correspond to basis states $\ket{K_{1,m}} \otimes \ket{K_{2,n}}$ and whose edges represent the actions of $H_1 \otimes I$ (horizontal) and $I \otimes H_2$ (vertical).  See Fig. (\ref{KrylovGrid}).

This grid can equivalently be reorganized into layers of constant $m+n$, 
\begin{align*}
m+n=0: &\quad (0,0)\\
m+n=1: &\quad (1,0), (0,1)\\
m+n=2: &\quad (2,0), (1,1), (0,2)\\
m+n=3: &\quad (3,0), (2,1), (1,2), (0,3)
\end{align*}
revealing a Pascal-triangle-like structure. The binomial coefficients governing the number of paths to each node encode how Krylov amplitudes distribute across these layers.   
\\ \\
As discussed, the Lanczos algorithm applied to the subsystems provide a decomposition
$$ H = (L_{1,+} + L_{2,+}) + (L_{1,0} + L_{2,0}) +  (L_{1,-} + L_{2,-})    $$
The repeated action of the Hamiltonian on states progress them along the diagonal layers of the grid through $(L_{1,+} + L_{2,+})$.  The action of $(L_{1,-} + L_{2,-})$ regresses a state back to a previous diagonal.  If the two subsystems do not evolve in synchrony, one does not regress to the original state but a linear combination of the original state and another orthogonal state in the same diagonal layer.  Similarly, the action of $(L_{1,0} + L_{2,0})$ can give rise to states in the same orthogonal layer.  Based on these considerations, we can define the \textbf{grid} complexity states in the following iterative way:  For each of the diagonal layers, we assign a complexity of $n$ if it first appears as a component of the state $H^n |\phi_r\rangle$.  Depending on the layer, there are other states on the same layer that are orthogonal to this state that will appear later in the Lanczos algorithm.  When this orthogonal state appears (say at order $m>n$) it is assigned a complexity of $m$ and so on.  The grid complexity has the same eigenvectors as $\hat{C}_1 + \hat{C}_2$ but the corresponding eigenvalues are always equal or larger.  It too, however, underestimates the combined complexity of the system.

\section{Continuum Limit}

The Krylov graph provides a discrete geometric description of operator growth under Hamiltonian evolution. In this section we show that, in an appropriate scaling limit, the dynamics of Krylov amplitudes admits a hydrodynamic description in terms of drift and diffusion on the product graph. This continuum limit clarifies the origin of superadditivity and provides a geometric interpretation of excess complexity as irreversible spreading in Krylov space.

\subsection{Discrete evolution on the product Krylov graph}
\noindent
Let $\psi_{m,n}(t)$ denote the amplitude of the product Krylov basis state
$\ket{K_{1,m}}\otimes\ket{K_{2,n}}$
in the time-evolved reference state $e^{-iHt}\ket{\phi_r}$, where
$H = H_1 \otimes I + I \otimes H_2$.
Each individual Krylov chain obeys a three-term Lanczos recursion relation,
\begin{equation}
H_i \ket{K_{i,m}} = b^{(i)}_{m+1}\ket{K_{i,m+1}} + a^{(i)}_m\ket{K_{i,m}} + b^{(i)}_m\ket{K_{i,m-1}}\,,
\label{eq:lanczos}
\end{equation}
with real Lanczos coefficients $a^{(i)}_m$ and $b^{(i)}_m>0$. Since the two Hamiltonians act independently on their respective subsystems, the Schrödinger equation for the product amplitudes factorises additively:
\begin{equation}
    i\,\partial_t \psi_{m,n} = (\mathcal{L}_x + \mathcal{L}_y)\psi_{m,n},
    \label{eq:discreteSE}
\end{equation}
where the lattice operators $\mathcal{L}_x$ and $\mathcal{L}_y$ generate nearest-neighbour hopping along the $m$ and $n$ directions of the Krylov graph,
\begin{align}
(\mathcal{L}_x\psi)_{m,n} &= b^{(1)}_{m+1}\psi_{m+1,n} + a^{(1)}_m\psi_{m,n} + b^{(1)}_m\psi_{m-1,n}\,, \\
(\mathcal{L}_y\psi)_{m,n} &= b^{(2)}_{n+1}\psi_{m,n+1} + a^{(2)}_n\psi_{m,n} + b^{(2)}_n\psi_{m,n-1}\,.
\end{align}
Equation \eqref{eq:discreteSE} has the form of a tight-binding Schrödinger equation on a two-dimensional lattice. The Krylov graph therefore plays the role of a discrete spacetime, with each Hamiltonian application corresponding to a single step along one of the lattice directions. The diagonals $m+n=\text{const}$ define shells of fixed total Krylov order, analogous to light-cone slices. Our interest lies not in the full phase-coherent dynamics of $\psi_{m,n}$, but in the coarse-grained spreading of probability on the Krylov graph, $P_{m,n}(t) = |\psi_{m,n}(t)|^2$. To obtain a closed hydrodynamic description for $P_{m,n}$ on the Krylov graph, we can proceed in either of two standard ways,
\begin{enumerate}
    \item Phase randomisation, via weak disorder or time averaging, which suppresses coherent current–current interference terms; or
    \item Semiclassical envelope approximation, in which $\psi_{m,n}$ is assumed to vary slowly compared to the lattice spacing, so that rapidly oscillating phase currents average to zero.
\end{enumerate}
Both procedures lead to the same effective description at leading order; probability evolves via local hopping on the lattice, governed by the Lanczos coefficients.\\
\noindent
 Focusing first on hopping in the $m$-direction, the corresponding master equation may be written as a one-dimensional birth--death process (with $n$ acting as a spectator),
\begin{equation}
    \partial_t P_{m,n}
    = b^{(1)}_{m} P_{m-1,n} +
    b^{(1)}_{m+1} P_{m+1,n} -
    \big(b^{(1)}_{m}+b^{(1)}_{m+1}\big) P_{m,n},
\label{eq:masterx}
\end{equation}
and similarly for hopping in the $n$-direction.  Equation \eqref{eq:masterx} expresses probability conservation under nearest-neighbour transitions induced by the Krylov recursion\footnote{A simple way to see this is that the sum over $m$ and $n$ vanishes.  Note also that the absence of the $a_m$ coefficients can be expected naturally since these can be absorbed as a phase into the $\psi_{m,n}$}.  This phase, of course, falls away when the probability is computed.  We emphasize that the above comes about from an approximation.  However, one may expect a similar equation to govern the general dynamics provided the coefficients are altered and (presumably small) couplings to next-to-nearest neig hbour couplings are included.   \\

\noindent
It is convenient to rewrite this equation in the form of a discrete continuity equation.
To this end, we define the probability current across the lattice edge
$(m,n)\to(m+1,n)$ as
\begin{equation}
    J^{(x)}_{m+1/2,n}
    \equiv
    b^{(1)}_{m+1} P_{m,n}
    -
    b^{(1)}_{m} P_{m+1,n}.
\label{eq:latticecurrent}
\end{equation}
A short calculation then shows that
\begin{equation}
    \partial_t P_{m,n}
    =
    -\nabla_x^- J^{(x)}_{m+1/2,n},
\end{equation}
and adding the analogous contribution from the $n$-direction yields a two-dimensional
lattice continuity equation. To make contact with hydrodynamics, we now perform a Kramers--Moyal expansion of the
one-step hopping process. Writing the current~\eqref{eq:latticecurrent} as
\begin{align}
J^{(x)}_{m+1/2,n}
&=
\frac12\!\left(b^{(1)}_{m+1}-b^{(1)}_{m}\right)
\big(P_{m,n}+P_{m+1,n}\big)
\nonumber\\
&\quad
-\frac12\!\left(b^{(1)}_{m+1}+b^{(1)}_{m}\right)
\big(P_{m+1,n}-P_{m,n}\big),
\end{align}
and retaining only the leading terms in a gradient expansion, we approximate
\(\tfrac12(P_{m,n}+P_{m+1,n})\simeq P_{m,n}\) and
\(P_{m+1,n}-P_{m,n}=(\nabla_x^+P)_{m,n}\).
This yields the drift--diffusion form
\begin{equation}
J^{(x)}_{m+1/2,n}
\approx
-\,v^{(1)}_m P_{m,n}
+
D^{(1)}_m \nabla_x^+ P_{m,n},
\end{equation}
with
\begin{equation}
v^{(1)}_m = b^{(1)}_{m+1}-b^{(1)}_{m},
\qquad
D^{(1)}_m = \tfrac12\big(b^{(1)}_{m+1}+b^{(1)}_{m}\big).
\end{equation}
An identical analysis applies in the $n$-direction. Substituting these expressions into
the lattice continuity equation gives,
\begin{align}
\partial_t P_{m,n}&=\nabla_x^-\!\Big[-v^{(1)}_m P_{m,n}+D^{(1)}_m\nabla_x^+P_{m,n}\Big]
+\nabla_y^-\!\Big[-v^{(2)}_n P_{m,n}+D^{(2)}_n\nabla_y^+P_{m,n}\Big],
\label{FP-eqn}
\end{align}
where $\nabla^\pm$ are forward/backward differences and
\begin{equation}\label{eq:rates}
v^{(i)}_m=b^{(i)}_{m+1}-b^{(i)}_{m},\quad 
D^{(i)}_m=\tfrac12\!\left(b^{(i)}_{m+1}+b^{(i)}_{m}\right).
\end{equation}

\subsection{Diffusive scaling and Fokker--Planck equation}
\noindent
We now pass from the discrete lattice description to a continuum (hydrodynamic)
description valid at large Krylov indices. To this end, we analytically continue the spatial coordinates on the Krylov graph,
\begin{equation}
    x=\varepsilon m,\quad y=\varepsilon n,
\end{equation}
and a mesoscopic time
\begin{equation}
    \tau=\varepsilon^2 t,
\end{equation}
with $\varepsilon\to 0$ and $x, y$ non-negative. This diffusive scaling is appropriate because the lattice
dynamics generated by \eqref{FP-eqn} is local and probability spreads over distances
$\Delta m,\Delta n\sim\sqrt{t}$. To proceed further, we assume that the Lanczos coefficients admit smooth envelopes,
\begin{equation}
   b^{(1)}_m \to b_1(x),\qquad b^{(2)}_n \to b_2(y),
\end{equation}
and similarly for $a^{(i)}$, and that the probability distribution varies smoothly
on the scale of the lattice spacing.\\

\noindent
Then, expanding the probability to second order in \(\varepsilon\) gives,
\begin{align}
    P_{m\pm1,n} &= P(x,y)\pm \varepsilon\,\partial_x P
    + \tfrac{\varepsilon^2}{2}\,\partial_x^2 P + \mathcal{O}(\varepsilon^3),\\
    P_{m,n\pm1} &= P(x,y)\pm \varepsilon\,\partial_y P
    + \tfrac{\varepsilon^2}{2}\,\partial_y^2 P + \mathcal{O}(\varepsilon^3)\,.
\end{align}
Inserting these expressions into the lattice continuity equation \eqref{FP-eqn}, we find
that the leading nontrivial contributions appear at order $\varepsilon^2$, as required
by the choice $\tau=\varepsilon^2 t$. Collecting terms and taking the limit $\varepsilon\to 0$, we find that the probability distribution
$P(x,y;\tau)$ obeys the two-dimensional Fokker--Planck equation
\begin{equation}
    \partial_\tau P
    =
    -\partial_x\!\big(v_1(x)P\big)
    -\partial_y\!\big(v_2(y)P\big)
    +
    \partial_x^2\!\big(D_1(x)P\big)
    +
    \partial_y^2\!\big(D_2(y)P\big),
    \label{eq:FPcont}
\end{equation}
with drift velocities and diffusion coefficients given by
\begin{equation}
   v_i(x)=\partial_x b_i(x),\quad D_i(x)=b_i(x).
\end{equation}
Several features of Eq.~\eqref{eq:FPcont} merit comment. First, no mixed
$\partial_x\partial_y$ term appears. This reflects the tensor-product structure of the
Hamiltonian; hops along the $x$ and $y$ directions originate from independent Krylov
chains and therefore remain uncorrelated at the level of local transport. Second, the diffusion coefficients are set directly by the Lanczos coefficients, establishing a direct link between Krylov geometry and hydrodynamic spreading. If the Lanczos coefficients saturate at large index, the dynamics reduces to simple diffusion,
whereas growing $b_i(x)$ generate a systematic drift toward larger Krylov order.
Finally, the diagonal Lanczos coefficients $a^{(i)}$ do not contribute at this order.
They enter the discrete evolution only through local phase rotations and act as a weak potential for the amplitudes; after phase averaging or in the envelope approximation, their effects drop out of the probability dynamics at leading hydrodynamic order.

\subsection{Constant-rate regime}

If we assume that the Lanczos coefficients approach constants, such that
\begin{equation}
    b_1(x) \to D_1, \quad b_2(y) \to D_2\,,
\end{equation}
the drift terms $v_i(x)=\partial_x b_i(x)$ vanish, and the Fokker--Planck
equation derived above reduces to a separable diffusion equation,
\begin{equation}
    \partial_\tau P(x,y;\tau)
    =
    D_1\,\partial_x^2 P
    +
    D_2\,\partial_y^2 P .
\end{equation}
This equation describes independent diffusion along the $x$ and $y$ directions of the
product Krylov graph. The fundamental solution, corresponding to a sharply localised
initial condition at the origin, is the anisotropic Gaussian
\begin{equation}
    P(x,y;\tau)
    =
    \frac{1}{\pi \tau \sqrt{D_1 D_2}}
    \exp\!\left[
    -\frac{x^2}{4D_1\tau}
    -\frac{y^2}{4D_2\tau}
    \right].
\end{equation}
Two important points about this solution deserve mention. First, the separability of the solution reflects the
tensor-product structure of the Hamiltonian; in the absence of drift, the two Krylov
directions remain dynamically independent. Second, the width of the distribution grows as
$\sqrt{\tau}$ in each direction, as expected for diffusive spreading.\\

\noindent
Although the diffusion is two-dimensional, complexity is controlled by the radial
coordinate corresponding to the total Krylov index. In the continuum description this
corresponds to the diagonal direction $x+y=\text{const}$, while in the discrete picture it corresponds to layers with fixed $m+n$. Projecting the Gaussian solution onto these diagonals reproduces the familiar binomial (Pascal-triangle) structure of the discrete Krylov graph. Concretely, the marginal distribution along the diagonal $x+y=s$ is itself Gaussian, with a width that grows as $\sqrt{s}$. This is precisely the continuum manifestation of the central limit theorem for the sum of two independent Bernoulli random walks. Each application of the Hamiltonian increments either $m$ or $n$, and after many steps the distribution of $m$ (at fixed $m+n$) becomes Gaussian. Thus, the central limit theorem controls the transverse spreading of probability within a
fixed Krylov shell. In the discrete problem, this explains why the binomial coefficients
governing the number of paths to a given node are sharply peaked near $m\simeq n$, and
why the distribution across each $m+n$ layer broadens as the square root of the layer
index.\\

\noindent
From the perspective of complexity, this observation is crucial. While the total
complexity depends only on the radial coordinate $m+n$, the orthogonal directions within each shell determine how much weight survives orthogonalisation into new Krylov basis vectors. The central limit theorem therefore provides the microscopic origin of the Gaussian envelopes and diffusive behaviour observed in the continuum theory.\\

\noindent
If, on the other hand, the Lanczos coefficients vary slowly with Krylov index, such that $b_i(x)=D_i+\delta b_i(x)$ with small gradients, the Fokker--Planck equation acquires drift terms,
\begin{align}
    \partial_\tau P
    &=
    -\partial_x\!\big(v_1(x)P\big)
    -\partial_y\!\big(v_2(y)P\big)
    \nonumber\\
    &\quad
    +\partial_x^2\!\big(D_1(x)P\big)
    +\partial_y^2\!\big(D_2(y)P\big),
\end{align}
with drift velocities $v_i(x)=\partial_x b_i(x)$. Physically, these drift terms bias
probability flow toward regions where the Lanczos coefficients grow. In Krylov space, this corresponds to a systematic tendency for operator weight to move toward larger Krylov
index whenever the local connectivity of the Krylov graph increases. In this regime, diffusion continues to control the transverse broadening across diagonal
shells, while advection controls the radial motion toward higher complexity. The
constant-rate case discussed above therefore represents a fixed point of the hydrodynamic
description, with weakly varying $b_i(x)$ providing controlled perturbations away from
purely diffusive behaviour.\\

\noindent
Taken together, the constant-rate and weak-drift regimes clarify the relationship between
the discrete combinatorics of the Krylov graph and the continuum hydrodynamic picture. The central limit theorem governs the Gaussian broadening within each fixed $m+n$ shell, while the Lanczos coefficients determine how probability flows between shells. This
separation of roles explains why excess complexity can remain bounded and parametrically small even as total complexity grows, and why tensor-product systems exhibit enhanced but non-chaotic operator spreading.

\subsection{Implications for (super)additivity}
\noindent
We can now use the hydrodynamic description to clarify the origin of superadditivity of
spread complexity. The spread (Krylov) complexity for the subsystems is given by
\begin{equation}
    C(t)=\sum_{m,n}(m+n)\,P_{m,n}(t),
\end{equation}
which in the continuum limit becomes
\begin{equation}
     C(\tau)=\int dx\,dy\,(x+y)\,P(x,y;\tau).
\end{equation}
Taking a time derivative and using the Fokker--Planck equation \eqref{FP-eqn}, we find
\begin{align}
    \frac{dC}{d\tau}
    &=
    \int dx\,dy\,(x+y)\,\partial_\tau P \nonumber\\
    &=
    \int dx\,dy\,(x+y)\Big[
    -\partial_x\!\big(v_1(x)P\big)
    -\partial_y\!\big(v_2(y)P\big)
    +\partial_x^2\!\big(D_1(x)P\big)
    +\partial_y^2\!\big(D_2(y)P\big)
    \Big].
\end{align}
Integrating by parts and assuming sufficiently rapid decay of $P$ at large
$x$ and $y$ as well as the boundary conditions at the origin, the diffusion terms drop out identically, while the drift terms
yield
\begin{equation}
    \frac{dC}{d\tau}
    =
    \int dx\,dy\,\big[v_1(x)+v_2(y)\big]\,P(x,y;\tau)
    =
    \langle v_1\rangle+\langle v_2\rangle .
\label{eq:Cgrowth}
\end{equation}
Several important conclusions follow immediately. First, diffusion does not contribute directly to the growth rate of complexity for the subsystems.  Diffusion controls the transverse spreading of probability within shells of constant $x+y$, but does not change the mean value of $x+y$ itself. This is a reflection of the fact that diffusive broadening redistributes weight within a fixed Krylov layer without pushing it radially outward.\\

\noindent
Second, the instantaneous growth rate of complexity is entirely controlled by the
drift velocities $v_i(x)=\partial_x b_i(x)$. Since the drift contributions from the
two Krylov directions add linearly, the growth rate takes the additive form shown in
\eqref{eq:Cgrowth}. This is the continuum counterpart of the additivity of drift
contributions in the discrete Krylov graph.
However, this does not imply that the total complexity is strictly additive.
When the Lanczos coefficients grow with Krylov index, the diffusion coefficients
$D_i(x)=b_i(x)$ also increase. This spatial growth of the diffusivity biases the
probability distribution toward larger values of $x+y$, effectively pushing weight
outward faster than in a pair of independent one-dimensional processes with fixed
coefficients. In the hydrodynamic picture, this manifests as an enhanced radial flux
induced by the curvature of the Krylov graph encoded in the growth of $b_i(x)$.\\

\noindent
This mechanism provides a transparent explanation of the strict inequality
\begin{equation}
    C_{12}(\tau)\ge C_1(\tau)+C_2(\tau)
\end{equation}
established in the discrete analysis. The inequality is saturated when the Lanczos
coefficients are constant, in which case the dynamics reduces to separable diffusion and
no additional radial bias is generated. In this constant-rate regime, the product
Krylov graph behaves as a direct sum of two independent chains, and complexity growth is
exactly additive. By contrast, whenever the Lanczos coefficients grow with Krylov index, the combined
system explores regions of the Krylov graph that are inaccessible to either subsystem
alone, leading to superadditive complexity growth. The hydrodynamic description thus
identifies spatial variation of the Lanczos coefficients---equivalently, curvature of
the Krylov graph---as the continuum origin of superadditivity.

\section{Conclusions and Outlook}
\noindent
In this work we have analysed the behaviour of Krylov complexity under tensor products of quantum systems, establishing a general and robust form of superadditivity. For Hamiltonians of the form
$H = H_1 \otimes I + I \otimes H_2$,
we proved that the corresponding complexity operator satisfies $C_{12} \;\ge\; C_1 + C_2$
for all target states supported on the Krylov subspace, with equality holding if and only if the two subsystems evolve synchronously. This result is entirely kinematic and fully general: it follows from the structure of the Lanczos algorithm and does not depend on integrability, chaos, or detailed spectral properties.\\

\noindent
A key conceptual advance of our analysis is the geometric interpretation of tensor-product Krylov dynamics. The tensor product naturally induces a higher-dimensional Krylov graph whose vertices correspond to tensor-product basis states and whose edges encode the independent actions of the subsystem Hamiltonians. Organising this graph into layers of constant total Krylov index reveals a Pascal-triangle–like structure, with binomial (and, more generally, multinomial) weights governing the distribution of amplitudes. The excess complexity is directly tied to deviations of the combined Krylov basis from this idealised layered structure, and can be interpreted as a measure of how strongly the Lanczos process mixes states within a given diagonal layer. Our explicit analysis of $su(2)\otimes su(2)$ examples illustrates these ideas concretely. In finite-dimensional settings the excess complexity is bounded and oscillatory, with its leading early-time behaviour scaling as either $t^4$ or $t^6$, depending on whether symmetry-enforced cancellations are present. These results clarify how superadditivity emerges dynamically, and why the combined complexity can remain close to additive for long times even when the inequality is strict.\\

\noindent
Taking a continuum limit, we showed that the Krylov amplitudes obey an effective advection–diffusion equation on the product graph. In this limit, the central-limit theorem explains the emergence of Gaussian envelopes across diagonal layers, while spatial variation of the Lanczos coefficients generates drift that pushes probability weight toward higher complexity. From this perspective, superadditivity acquires a simple physical interpretation; curvature in the effective Krylov geometry enhances operator growth beyond the sum of independent one-dimensional processes.\\

\noindent
From the perspective of quantum chaos and scrambling, our results clarify an important distinction. Although tensor products of integrable Hamiltonians will again be integrable, their Krylov geometry is nonetheless nontrivial. Superadditive spread complexity demonstrates that operator growth can become more intricate purely through combinatorial mixing of operator pathways, without invoking chaos in the traditional sense. In this way, tensor-product systems provide a controlled arena in which to separate the roles of entanglement, interaction, and Krylov connectivity in shaping complexity growth. Our results therefore complement recent work relating Krylov complexity to chaos diagnostics, OTOCs, and spectral statistics, by highlighting a regime where enhanced complexity arises independently of scrambling in the usual sense.\\

\noindent
Several natural extensions of this work suggest themselves: 

\begin{itemize}
    \item First, two technical generalization should be readily possible.  We have focused on the case of complexity i.e. the first moment of the distribution of the Krylov probabilities, but the proof should hold for higher moments.  Also, a recent conjecture that the total K- or spread complexity of a system is greater than their symmetry resolved versions \cite{Caputa:2025mii, Caputa:2025ozd}.  We suspect that our proof for the positivity of excess complexity can be adjusted to prove this interesting conjecture as well.  
    \item Second, while we focused primarily on two-factor tensor products, the framework generalizes straightforwardly to \textbf{multi-partite systems}. In this case the Krylov graph becomes an $N$-dimensional lattice, with constant-complexity layers forming $(N-1)$-simplices. Understanding how superadditivity scales with the number of subsystems, and whether new collective effects emerge in large-$N$ limits, would be particularly interesting.
    \item Third, our results highlight a sharp distinction between tensor-product algebras such as $su(2)\oplus su(2)$ and genuinely \textbf{non-factorising algebras} such as $su(4)$. Incorporating entangling generators modifies the Krylov graph connectivity itself, rather than merely its weighting, and is expected to produce qualitatively stronger complexity growth. A systematic comparison between tensor-product and simple Lie algebras could shed light on the role of entanglement generation in operator growth and complexity.
    \item Fourth, it would be valuable to connect the \textbf{Krylov-graph picture} developed here with other notions of quantum complexity, including circuit complexity and Nielsen geometry. In particular, the excess complexity identified here may provide a diagnostic for how much ``nonlocal structure is required to simulate tensor-product dynamics using restricted gate sets.  This may be especially relevant to probes of the Mpemba effect using variations of Krylov complexity \cite{Beetar:2025tlf, Alishahiha:2025sep}. \item Finally, from a broader perspective, the emergence of \textbf{effective diffusion} equations and geometric structures in Krylov space reinforces the idea that operator growth admits a coarse-grained, hydrodynamic description. Exploring the universality of this description—especially in chaotic many-body systems, periodically driven systems, or systems with constraints such as symmetries or locality—remains an open and promising direction.
\end{itemize}

\noindent
Taken together, these results position the Krylov graph as a unifying geometric object underlying operator growth, complexity, and tensor-product structure in quantum systems, and suggest that much remains to be learned by studying its higher-dimensional and non-factorising generalisations.

\section*{Acknowledgements}
JM and HJRVZ
are supported in part by the ``Quantum Technologies for Sustainable Development" grant
from the National Institute for Theoretical and Computational Sciences of South Africa
(NITHECS).

\bibliographystyle{JHEP}
\bibliography{biblio.bib}

\end{document}